\documentclass{PoS}

\PoS{PoS(LAT2005)296}

\title{Gauge invariant 'monopoles' and color confinement mechanism}

\ShortTitle{Gauge invariant 'monopoles' and color confinement mechanism}

\author{\speaker{Tsuneo Suzuki}\\
        Institute for Theoretical Physics, Kanazawa University, Kanazawa
        920-1192, Japan\\
        and RIKEN, Radiation Laboratory, Wako 351-0158, Japan\\
        E-mail: \email{suzuki@hep.s.kanazawa-u.ac.jp}}

\author{Katsuya Ishiguro, Yoshifumi Nakamura and Toru Sekido\\
        Institute for Theoretical Physics, Kanazawa University, Kanazawa
        920-1192, Japan \\
        and RIKEN, Radiation Laboratory, Wako 351-0158, Japan}

\abstract{
The  dual Meissner effect is described and numerically
observed in a gauge-invariant way in lattice Monte-Carlo simulations of
pure $SU(2)$ QCD.
A gauge-invariant monopole-like quantity on the lattice is defined by
a gauge-invariant Abelian-like field strength. 
The  Abelian-like field strength is expressed  in terms of a
 unit-vector in color space
which is constructed by a non-Abelian field strength itself. It is just equal to 
the absolute value of the corresponding non-Abelian field strength except for the sign.
In this note we show the theoretical background and most numerical results will be 
published in a separate report~\cite{Suzuki:2005lat052} in this conference. 
}

\FullConference{XXIIIrd International Symposium on Lattice Field Theory\\
		 25-30 July 2005\\
		 Trinity College, Dublin, Ireland}

\newcommand{\beqn}{\begin{eqnarray}}
\newcommand{\eeqn}{\end{eqnarray}}

\begin{document}

\section{Introduction}

One of the most essential problems of color confinement in QCD is to explain the mechanism of the flux squeezing of 
non-Abelian electric fields between a pair of static quark and antiquark.  In $SU(2)$ QCD, $(E^a_i)^2$ or $\sqrt{(E^a_i)^2}$ is expected to be squeezed to reproduce the linear static potential. Numerically the expected squeezing of the gauge-invariant combination of the electric field was observed beautifully in  lattice 
$SU(2)$ QCD~\cite{Bali:1994de}.

Thirty years ago, 'tHooft~\cite{tHooft:1975pu} and Mandelstam~\cite{Mandelstam:1974pi} conjectured that the dual Meissner effect is  the color confinement mechanism of QCD. 
However what causes the dual Meissner effect and how to treat the non-Abelian property were not clarified.
An interesting idea is to utilize a topological monopole like the 'tHooft-Polyakov monopole~\cite{'tHooft:1974qc, Polyakov:1974ek} found in $SU(2)$ QCD with an adjoint Higgs field $\phi$. A topological monopole has a bare magnetic charge satisfying the Dirac quantization condition with a bare electric charge. 
An important quantity is a 'tHooft Abelian-like field strength 
\begin{eqnarray}
f_{\mu\nu}&=&n^aF^a_{\mu\nu}+\epsilon_{abc}n^a(D_{\mu}n)^b(D_{\nu}n)^c, \label{thooft-fmunu}
\end{eqnarray}
where $n^a$ is a unit vector in color space transforming as an adjoint representation
and $(D_{\mu}n)^b$ is a covariant derivative. 
In $SU(2)$ QCD with a Higgs field, $\hat{\phi}^a$ is adopted as $n^a$, since there is a 
classical monopole solution corresponding to the choice. 

However one may choose any 
adjoint operator for $n^a$ to discuss a topological monopole~\cite{tHooft:1981ht} in quantum field theory.
This observation is important in real QCD without a Higgs field. We can discuss 
a topological monopole in terms of $n^a$ which is constructed in terms of gluon fields.
A monopole picture can be seen more clearly if we 
project further $SU(3)$ QCD to an Abelian $U(1)^2$ theory by a partial gauge fixing~\cite{tHooft:1981ht}. After the Abelian projection, we have an Abelian $U(1)^2$ theory with 
Abelian electric and magnetic charges. It is conjectured in Ref.~\cite{tHooft:1981ht}
that the condensation of the Abelian monopoles causes the dual Meissner effect explaining the color confinement. However there is a serious problem in this scenario.
Namely there exist infinite ways of choosing $n^a$ or in other words infinite possible Abelian projections. Moreover, the monopole condensation, if happens, can explain only the squeezing of an Abelian-like electric field $f_{0i}$ defined in Eq.(\ref{thooft-fmunu}). How good an approximation it is to the real and expected flux squeezing of $\sqrt{(E^a_i)^2}$ depends strongly on the choice of $n^a$. 

An Abelian projection adopting a special gauge called Maximally Abelian gauge (MA)~\cite{Suzuki:1983cg,Kronfeld:1987ri,Kronfeld:1987vd} is found to give us interesting results~\cite{Suzuki:1992rw,Chernodub:1997ay,Suzuki:1998hc}  supporting 
importance  of the Abelian monopoles. In this case, the Abelian electric field 
can approximate very well the long-range behavior of the non-Abelian one, since other components are suppressed. However such beautiful results are not seen in other general gauges than the MA gauge.

It is the purpose of  this note and the separate report~\cite{Suzuki:2005lat052} to show numerically
that the  dual Meissner effect occurs in a gauge-invariant way with the use of  a  gauge-invariant Abelian-like field strength and a monopole-like quantity. 
We do not need  any 
Abelian projection nor any gauge-fixing.
In this note we explain the theoretical background of our idea and 
show most of numerical results in the separate report~\cite{Suzuki:2005lat052}.

\section{Abelian-like field strength}
We define an Abelian-like field strength:
\begin{eqnarray}
f_{\mu\nu}(x)&=&\vec{n}_{\mu\nu}(x)\cdot \vec{F}_{\mu\nu}(x), \label{fmunu}
\end{eqnarray}
where the summation  over $\mu$ and $\nu$ is not taken\footnote{We used a little different Abelian-like field strength with an additional term as in 'tHooft field strength. However the additional term is not essential in the following discussions, since we are not 
dealing with a topological monopole. Hence we adopt the above definition (\ref{fmunu}) for simplicity.}.  $\vec{n}_{\mu\nu}$ is a unit vector in color space transforming as an
adjoint representation in SU(2). Note that
$f_{\mu\nu}$ is not a simple Lorentz tensor. 

 Explicitly we adopt the following unit vector in color space of $SU(2)$ QCD~\cite{Chernodub:2000wk,Chernodub:2000bq,Chernodub:2000rg}:
\begin{eqnarray}
n^a_{\mu\nu}(x)&=&\epsilon_{\mu\nu}\frac{F^a_{\mu\nu}(x)}{\sqrt{\sum_{a=1}^3(F^a_{\mu\nu}(x))^2}}, \label{eq.SU2}
\end{eqnarray}
where $\epsilon_{\mu\nu}$ is an antisymmetric tensor with the sign
convention $\epsilon_{\mu<\nu}=1$. The opposite sign convention can be adopted which means the existence of the sign ambiguity. The continuity could determine the relative sign. 
$F^a_{\mu\nu}$ is a non-Abelian field strength with a color charge $a$ and no summation is taken with respect to $\mu$ and $\nu$ in Eq.(\ref{eq.SU2}). This 
choice is unique in a sense that  Eq.(\ref{fmunu})
is just equal to the gauge-invariant absolute value of the non-Abelian
field strength itself except for the sign in $SU(2)$ QCD.
Actually an electric field component $E_i$ defined by $f_{4i}$ is
$-\sqrt{(E^a_i)^2}$ the squeezing of which is to be explained.

A gauge-invariant monopole-like quantity is defined from the violation
of the Bianchi identity:
\begin{eqnarray}
k_{\mu}(x)=\frac{1}{8\pi}\epsilon_{\mu\nu\alpha\beta}\partial_{\nu}
f_{\alpha\beta}(x). \label{monopole}
\end{eqnarray}
This is conserved but is not a simple Lorentz vector.
 Hereafter we call the monopole-like quantity simply as 'monopole'.
We get from Eq.(\ref{monopole})
\begin{eqnarray}
\vec{\nabla}\times\vec{E}+\partial_{4}\vec{B}&=&4\pi\vec{k}, \label{BI}\\
\vec{\nabla}\cdot\vec{B}&=&-4\pi k_4, \label{BI-2}
\end{eqnarray}
where
\begin{eqnarray}
\vec{E}&\equiv&\left(-\sqrt{(E^a_1)^2} , -\sqrt{(E^a_2)^2} , -\sqrt{(E^a_3)^2}\right), \\
\vec{B}&\equiv&\left(\sqrt{(B^a_1)^2}, -\sqrt{(B^a_2)^2}, \sqrt{(B^a_3)^2}\right).
\end{eqnarray}
Note that the magnetic charge 
 defined in Eq.(\ref{BI-2}) does not satisfy the Dirac quantization condition with respect to  bare charges contrary to the usual case of a magnetic charge defined in terms of 
a 'tHooft field strength.

\section{'Monopole'  on the lattice}

Now we go to a lattice QCD framework.
A non-Abelian field strength $F_{\mu\nu}(s)$ is 
given by a $1\times 1$ plaquette variable defined by 
a path-ordered product of four non-Abelian link matrices on the lattice:
\begin{eqnarray*}
U_{\mu\nu}(s)= \exp\left(iF_{\mu\nu}(s)\right)= U^0_{\mu\nu}(s)+iU^a_{\mu\nu}(s)\sigma^a.
\end{eqnarray*}
The unit vector in color space is 
\begin{eqnarray}
n^a_{\mu\nu}(s) = \epsilon^{\mu\nu}{U^a_{\mu\nu}(s)\over \sqrt{1-(U_{\mu\nu}^0(s))^2}},\label{unit-vector}
\end{eqnarray}
and the Abelian-like field strength is written similarly as 
in Eq.(\ref{fmunu})
\begin{eqnarray}
f_{\mu\nu}(s)&=&n^a_{\mu\nu}(s)F^a_{\mu\nu}(s) .
\end{eqnarray}
This definition is explicitly gauge-invariant.

We define a gauge-invariant lattice 'monopole' in the same way as in Eq.(\ref{monopole}):
\begin{eqnarray}
k_{\mu}(s)&=&\frac{1}{8\pi}\epsilon_{\mu\nu\alpha\beta}\Delta_{\nu}
f_{\alpha\beta}(s+\hat{\mu}), \label{lattice-monopole1}
\end{eqnarray}
which satisfies $\Delta'_{\mu}k_{\mu}(s)=0$.  $\Delta_{\mu}$
($\Delta'_{\mu}$) is a lattice forward (backward) derivative.
Note that this 'monopole' is gauge-invariant and conserved 
but is not integer. In the separate report~\cite{Suzuki:2005lat052} we
will see that the electric field $\sqrt{(E^a_z)^2}$ between a static
quark pair is actually squeezed due to the solenoidal 'monopole
current'. The dual Meissner effect can be seen in a gauge-invariant way. 

\section{Comparison with Abelian monopoles after  Abelian projections}
Here we compare our gauge-invariant 'monopole' with 
a (topological) monopole after an Abelian 
projection. 
In the latter case, an Abelian link variable $\theta_{\mu}^{AP}(s)$ is defined by a phase 
of the diagonal part of a non-Abelian link field after a gauge fixing.
An Abelian field strength $\theta_{\mu\nu}^{AP}(s)$ is defined as
$\theta_{\mu\nu}^{AP}(s)\equiv\theta_{\mu}^{AP}(s)+\theta^{AP}_{\nu}(s+\hat{\mu})-\theta_{\mu}^{AP}(s+\hat{\nu})-\theta^{AP}_{\nu}(s).$
An Abelian 
monopole is defined as~\cite{DeGrand:1980eq}
\begin{eqnarray}
k^{AP}_{\mu}(s)={1\over 8\pi}\epsilon_{\mu\nu\alpha\beta}\Delta_{\nu}
\theta^{AP}_{\alpha\beta}(s+\hat{\mu}).
\end{eqnarray}
This is conserved and takes an integer number. Namely it is a topological monopole. It is known that, if we perform the MA gauge fixing where $\sum_{s,\mu} Tr [ U_{\mu}(s)\sigma_3U_{\mu}^{\dagger}(s)\sigma_3]$
is maximized, we get interesting results called as monopole dominance~\cite{Suzuki:1992rw,Chernodub:1997ay,Suzuki:1998hc}. However, if we adopt a gauge-fixing diagonalizing $F_{12}(s)$ ($F_{12}$ gauge), 
such interesting results are not seen~\cite{Suzuki:1992rw}. The $F_{12}$ gauge fixing on the lattice is defined in such a way as 
$U_1(s)U_2(s+\hat{1})U_1^{\dagger}(s+\hat{2})U_2^{\dagger}(s)$ is
diagonalized.

\begin{figure}[htb]
\begin{center}
\includegraphics[height=5cm]{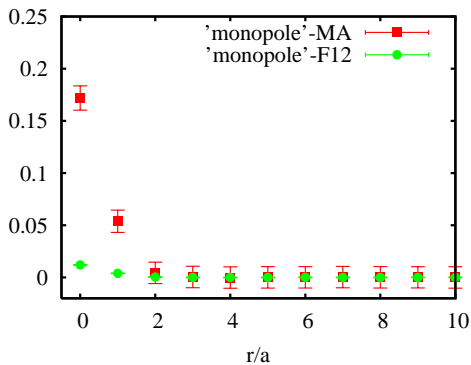}
\caption{\label{corr-mon}Correlation of the 'monopole' with Abelian
 projected 
monopoles in MA gauge and F12 gauge.}
\end{center}
\end{figure}

We show  numerical results of
 the correlations $C(r)$ between the 'monopole'
and Abelian projected monopoles in Fig.\ref{corr-mon}.
%
%
%
\begin{eqnarray}
C(r)&=&{\langle \vert k^{AP}_{\mu}(0) \vert \vert k'_{\mu}(r) \vert \rangle \over 
\langle \vert k^{AP}_{\mu}(0) \vert \rangle \langle \vert k'_{\mu}(0)
\vert \rangle }
-1.
\end{eqnarray}
The above two kinds of  Abelian monopoles are investigated.
One of them is a monopole in the MA gauge fixing.
It is interesting that the correlation  is very strong.
On the other hand the correlation is very weak between the gauge-invariant 'monopole' and an Abelian  monopole in the $F_{12}$ gauge. 

Note that the Abelian-like field strength Eq.(\ref{fmunu}) is reduced to
an Abelian one if off-diagonal components are negligible. 
This occurs in 
the MA gauge 
and in the  maximally Abelian Wilson loop gauge~\cite{Shoji:1999gj} where almost the
same fine results as in the MA gauge are observed. 
In these cases, we can adopt $\vec{n}_{\mu\nu}=(0,0,1)$, 
since only the diagonal component exists. Hence the  gauge-invariant results we are going to show in the report~\cite{Suzuki:2005lat052} could explain why only restricted Abelian projection schemes like the MA gauge look nice among infinite possible candidates. In the case of the $F_{12}$ gauge, such a reduction does not occur.

\section{Conclusion}
We have defined a gauge-invariant monopole-like
quantity by using an Abelian-like field strength on the lattice.
This current is not a simple Lorentz vector and does not take an integer number.
We have compared the gauge-invariant 'monopole' with Abelian topological monopoles
appearing after Abelian projections. A strong correlation is observed between 
the gauge-invariant monopole and the Abelian monopole in the MA gauge, whereas 
no correlation is seen with the Abelian monopole in $F_{12}$ gauge. When the unit-vector $\vec{n}_{\mu\nu}$ in color space is well approximated as  $\vec{n}_{\mu\nu}=(0,0,1)$, our gauge-invariant Abelian-like field strength becomes Abelian. Such a situation is expected in the MA gauge and the maximally Abelian Wilson loop gauge~\cite{Shoji:1999gj} where dominance of the topological Abelian monopole is seen.

In Ref.~\cite{Suzuki:2005lat052} we will show numerically
that the  dual Meissner effect occurs in a gauge-invariant way
 with the use of a gauge-invariant Abelian-like field strength and
'monopoles'.
 We do not need any Abelian projection nor any gauge-fixing.

\begin{acknowledgments}
The numerical simulations of this work were done using RSCC computer clusters in 
RIKEN. The authors would like to thank RIKEN for their support of computer facilities. 
T.S. is supported by JSPS Grant-in-Aid for Scientific Research on Priority Areas 13135210 and (B) 15340073.
\end{acknowledgments}


\end{document}